# Structural and Optical Properties of Silicon Nanocrystals Embedded in Silicon Carbide: Comparison of Single Layers and Multilayer Structures

Charlotte Weiss[1], Manuel Schnabel[1], Andreas Reichert[1], Philipp Löper[1,2], Stefan Janz[1]

[1]*Fraunhofer Institute for Solar Energy Systems, Heidenhofstraße 2, 79110 Germany.*

[2]Now with*: École Polytechnique Fédérale de Lausanne, Institute of Microengineering, Photovoltaics and Thin Film Electronics Laboratory, Rue de la Maladière 71b, CH-2002 Neuchâtel 2, Switzerland.*

0 ABSTRACT

The outstanding demonstration of quantum confinement in Si nanocrystals (Si NC) in a SiC matrix requires the fabrication of Si NC with a narrow size distribution. It is understood without controversy that this fabrication is a difficult exercise and that a multilayer (ML) structure is suitable for such fabrication only in a narrow parameter range. This parameter range is sought by varying both the stoichiometric SiC barrier thickness and the Si-rich SiC well thickness between 3 nm and 9 nm and comparing them to single layers (SL).

The samples processed for this investigation were deposited by plasma-enhanced chemical vapor deposition (PECVD) and subsequently subjected to thermal annealing at 1000-1100°C for crystal formation.

Bulk information about the entire sample area and depth were obtained by structural and optical characterization methods: information about the mean Si NC size was determined from grazing incidence x-ray diffraction (GIXRD) measurements. Fourier-transform infrared spectroscopy (FTIR) was applied to gain insight into the structure of the Si-C network, and spectrophotometry measurements were performed to investigate the absorption coefficient and to estimate the bandgap $E_{04}$.

All measurements showed that the influence of the ML structure on the Si NC size, on the Si-C network and on the absorption properties is subordinate to the influence of the overall Si content in the samples, which we identified as the key parameter for the structural and optical properties. We attribute this behavior to interdiffusion of the barrier and well layers. Because the produced Si NC are within the target size range of 2-4 nm for all layer thickness variations, we propose to use the Si content to adjust the Si NC size in future experiments.

1 INTRODUCTION

In the field of photovoltaic research and development, much work is devoted to efficiency enhancement and cost reduction. One promising route to address both issues is the development of tandem solar cells that consist of two or more stacked single cells. These single cells must exhibit decreasing bandgaps from the illuminated side to the rear side to convert different wavelength ranges of the solar spectrum, minimizing thermalization and transmission losses [1]. For cost reduction and industrial acceptance, it is advantageous to focus on the established Si technology for tandem solar cell concepts. The ideal bandgap for a top cell on a crystalline Si (1.1 eV) bottom cell was calculated to be 1.7 eV [2]. There are two ways to produce Si solar cells with a larger bandgap than crystalline Si. The first possibility is to use hydrogenated amorphous Si (a-Si:H) [3]. This material is already widely used in Si tandem photovoltaics [4, 5] but suffers from light-induced degradation [1, 6]. The second possibility is to use the effect of quantum confinement [7], which means enlarging the Si bandgap by the reduction of the Si crystals to the nanocrystal (NC) scale. Not only the size of the NC but also the band structure of the embedding matrix material and the interdot distance determine the properties of the Si NC [8]. Several groups use 3C-SiC as a matrix material, as it provides a small conduction band offset of 0.5 eV compared with other matrix materials (1.9 eV for $Si_3N_4$ or 3.2 eV for $SiO_2$ [7]), with the same trend observed for the valence band. A small band offset increases the tunneling probability from one Si NC to the other and hence the conductivity of the material, making transport less sensitive to variations in NC separation [7, 8]. Although material quality has improved in recent years [9-16], the recently presented first all-Si tandem solar cell with a Si NC top cell absorber shows low efficiency and no clear evidence of quantum confinement [17]. Therefore, further investigation of Si NC in SiC is necessary and is the topic of this work.

Löper *et al.* [18] predicted that for Si NC with an interdot distance of 2 nm in a SiC matrix, the ideal bandgap of 1.7 eV corresponds to a Si NC size of 2-3 nm. This target Si NC size is assumed to be valid for a wide interdot range due to simulations of Jiang *et al.* [8] which show an influence of the interdot distance of Si NC in SiC mainly on the bandwidth and less on the bandgap. To reach Si NC size control for the adjustment of the Si bandgap, the so-called multilayer (ML) approach has been developed (see Figure 1).

Alternating layers of stoichiometric SiC barrier layer (SiC) and Si rich SiC well layer (SRC) with thicknesses in the range of 1-10 nm are deposited. During the subsequent annealing step, phase separation and Si NC formation are expected to occur in the SRC sublayers, while the SiC sublayer should serve as a diffusion barrier. The ML approach is known to work very well for Si NC size control in $SiO_2$ [19-21] but is much more challenging in SiC because interdiffusion of the SiC barrier layer and the SRC well layer occurs [12]. In addition, co-crystallization of Si and SiC NC is observed [22-26].

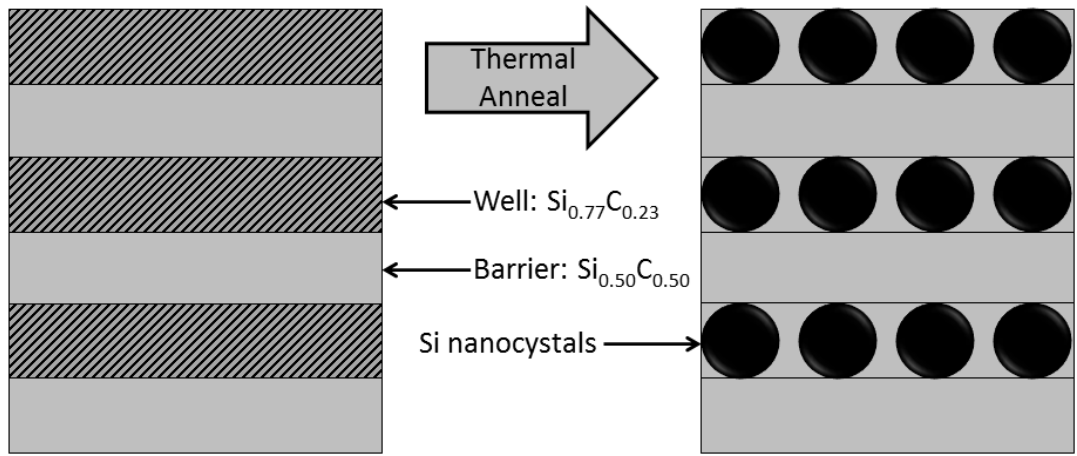

**Figure 1: The multilayer approach: A stack of SiC/SRC bilayers are deposited and annealed. Ideally, SRC sublayer thickness controls Si NC size, while the SiC sublayer thickness has no effect on crystallization.**

It was only recently that Summonte *et al.* [12] managed to obtain size-controlled Si NC in SiC by optimizing the ML parameters: they proved using transmission electron microscope (TEM) images that for an as-deposited SRC thickness between 3 and 4 nm, the ML structure survives the annealing for both 3 nm and 9 nm thick SiC barrier layers. In the case of the 3 nm barrier layer, no size controlled Si NC were achieved, but either a continuous crystallized well (for 4 nm SRC layer) or an outgrowth of the Si NC in the barrier layers was observed (for 3 nm SRC layer). For the thicker barrier of 9 nm, both outgrowth and continuous crystallization were less pronounced, but it could still not be excluded that the Si NC touch each other. An interconnection of the Si NC seems probable because the Si content in the SRC layers is quite high (x = 0.85). However, the authors claim that separated, ordered Si NC should be achieved for a certain combinations of Si content and thickness of the SRC layer.

In this work, we study the effect of sublayer thicknesses on Si NC morphology over a wider parameter range by varying the SiC barrier thickness and the SRC well thickness between 3 nm and 9 nm, with a Si content of 77% in the well. For comparison, single layers (SL) with different Si contents were investigated. We determined the mean NC size by grazing incidence x-ray diffraction (GIXRD). To achieve a deeper understanding of the phase transformations in the layers, we conducted a detailed investigation of the Si-C network by Fourier transformed infrared spectroscopy (FTIR). Finally, we tried to identify signs of quantum confinement by determining the absorption behavior and the bandgap $E_{04}$ from $R\&T$ measurements in the UV-Vis wavelength range. The advantage of all mentioned characterization methods is, that they yield averaged information over a macroscopic sample area and depth.

2 EXPERIMENTAL

## 2.1 Sample Preparation

Two types of substrates were used for layer deposition: 250 μm thick p-type FZ silicon, (100)-oriented with a resistivity of 10 Ωcm for GIXRD and FTIR measurements, and 1 mm thick fused silica (Suprasil®1) for spectrophotometry measurements. Both types of substrates were cleaned in hot $HNO_3$ and dilute HF before film deposition. The fused silica was subjected to an additional cleaning step of hot $HCl/H_2O_2$ solution followed by a second HF etching step before film deposition.

All ML and SL used in this work were deposited by plasma-enhanced chemical vapor deposition (PECVD) conducted in a Roth&Rau AK400 reactor. The pressure during deposition was kept at 0.3 mbar and the substrate temperature at 270°C. The plasma frequency was 13.56 MHz, and the plasma power density was 100 mW/cm².

The variation of the gas fluxes of $SiH_4$, $CH_4$ and $H_2$ allows the deposition of hydrogenated amorphous silicon carbide (a-$Si_xC_{1-x}$:H) with varying stoichiometry. The three different layer compositions used in this work were determined by Rutherford Backscattering Spectrometry with an accuracy of 1% [27] and are listed in Table 1 with the associated gas fluxes.

**Table 1: Reactant gas fluxes used for PECVD in sccm.**

|  | $SiH_4$ | $CH_4$ | $H_2$ |
|---|---|---|---|
| stoichiometric SiC a-$Si_{0.50}C_{0.50}$:H | 7 | 60 | 100 |
| Si rich SiC 1 a-$Si_{0.63}C_{0.37}$:H | 7 | 20 | 100 |
| Si rich SiC 2 a-$Si_{0.77}C_{0.23}$:H | 10 | 10 | 100 |

The deposited ML consist of 20 bilayers of alternating a-$Si_{0.50}C_{0.50}$:H (SiC) and a-$Si_{0.77}C_{0.23}$:H (SRC). The thickness of these sublayers was varied between 3 – 9 nm, resulting in total as-deposited layer thicknesses of 187 – 334 nm. Three different types of SL of approximately 200 nm thickness were deposited: SL with the SiC and SRC sublayer composition ($Si_{0.50}C_{0.50}$ and $Si_{0.77}C_{0.23}$) and SL with a composition of $Si_{0.63}C_{0.37}$, which corresponds to the overall Si content of a ML with SRC/SiC sublayer thicknesses of 9 nm/6 nm.

It is important to note that during this work, it was not possible to verify the overall Si content in the ML experimentally. Because the ML deposition by PECVD consists of 40 individual layers, it is most likely that the frequent change of gas flow and the deposition conditions at interfaces influence the incorporated amount of Si.

After deposition, an annealing step at 1000°C for 60 min or at 1100°C for 30 min was conducted in a conventional quartz tube furnace under $N_2$ (99,9999%) atmosphere. The ramping up was performed with 10°C/min under a $N_2$ flux of 10 slm. Between 300°C and 800°C, H-effusion occurs [11], resulting in a reduction of film thickness between 20% and 30%. From approximately 900°C on, solid phase separation in the SRC layers and the formation of Si NC begins. This Si crystallization occurs together with the formation of c-SiC in the whole sample [28].

## 2.2 Sample Characterization

***GIXRD*** patterns were recorded using a Philips X'Pert MRD system equipped with a CuK$_\alpha$ x-ray ($\lambda$ = 0.154 nm) source. The angle of incidence was set to $\omega$ = 0.3° (for maximum intensity), and 2$\theta$ was varied between 10° and 90°. All patterns were background-corrected by subtracting an exponential decay function. The Si(111) peak at 28.4° and the SiC(111) peak at 35.7° were fitted by a Lorentz function, and their full width at half maximum *FWHM* was used to determine the mean grain size *L* of the Si and SiC NC by means of the Scherrer equation:

$$L = \frac{K \cdot \lambda}{FWHM \cdot cos\theta} \quad (1)$$

where $\theta$ is the diffraction angle of the analyzed reflection, $\lambda$ is the x-ray wavelength and *K* is a form factor, which is 0.9 for spherical crystals in the cubic crystal system [22, 29]. It should be emphasized that this grain size determination does not provide information on the grain size distribution and is therefore referred to as an estimation of the mean grain size. Additionally, the fitting process gives rise to an error of ± 0.5 nm in grain size. However, the formation of large grains (> 10 nm) can by excluded for all our samples, as we know that their formation causes the appearance of a second Lorentzian shaped GIXRD reflex that interferes with the reflex belonging to smaller grains.

***FTIR*** spectroscopy was conducted in the range of 400 cm$^{-1}$ to 4000 cm$^{-1}$ with 6 cm$^{-1}$ resolution using a Bruker IFS 113v instrument on layers processed on Si substrates. To remove the absorbance signal of the Si substrate, a reference substrate was measured under identical optical conditions, and its signal was subtracted from the absorbance of all samples. Then, the absorption coefficient ($\alpha(\nu)$) spectra were calculated from the measured absorbance $B(\nu)$ with the help of the sample thickness *d* by the following relation:

$$\alpha(\nu) = ln10 \frac{B(\nu)}{d} \quad (2)$$

The main mode of all FTIR spectra is the Si-C stretching vibration at approximately 800 cm$^{-1}$ ([30-32] and references therein). To gain further information from FTIR measurements beyond the identification of the vibration modes, accurate peak fitting and a careful interpretation of the results is needed. The FTIR peak fitting was performed by the following routine. First the background was fitted by a higher-order polynomial and subtracted manually. Because there is, to the authors' knowledge, no analytical approach to FTIR background correction, this process was deemed more reliable than an automated background correction routine. As a second step, the Si-C mode approximately 800 cm$^{-1}$ [30] was fitted by a combination of a Lorentz $L(v)$ and a Gauss $G(v)$ peak, as it is known that the Si-C network after annealing consists of both crystalline and amorphous domains, and the Lorentz part of the peak area can be attributed to the crystalline Si-C phase in the samples while the Gaussian part arises from a random and hence amorphous Si-C bond distribution [33-36]. Thus, the fit function can be written as

$$\alpha(v) = L(v) + G(v) \quad (3)$$

$$L(v) = A_L \cdot \frac{2}{\pi} \cdot \frac{w_L}{4(v - v_L)^2 + w_L^2} \quad (4)$$

$$G(v) = A_G \cdot \frac{1}{w_G} \sqrt{\frac{4\ln(2)}{\pi}} \cdot exp\left[-4\ln(2) \cdot \left(\frac{v - v_G}{w_G}\right)^2\right] \quad (5)$$

where $A_L$ ($A_G$) is the area of the peak, $v_L$ ($v_G$) is the peak position, and $w_L$ ($w_G$) is the FWHM of the Lorentz (Gaussian) peak. The fitting parameters give the following information about the network in the samples.

The **peak position $v_0$** indicates which vibration modes and hence which bonds are present in the network. In contrast to FTIR analysis of gases, there are no free vibrating molecules in solid-state samples, which is why the surrounding network influences every vibration mode. As a result, every change in the bonding network will be detected by a shift of the peak position of the Si-C vibration mode. The **peak area $A$** is proportional to the bond density of the corresponding absorption mode [37]:

$$N(Si - C) = Q_{Si-C} A = Q_{Si-C} \int_{v_1}^{v_2} \alpha(v) dv. \quad (6)$$

The proportionality constant $Q_i$ can be determined experimentally. In the literature, the integration over $\frac{\alpha(\nu)}{\nu}$ and the determination of the proportionality constant $K_{Si-C}$ constitute a more common approach, giving the following:

$$N(Si-C) = K_{Si-C} \int_{\nu_1}^{\nu_2} \frac{\alpha(\nu)}{\nu} d\nu. \quad (7)$$

Based on a given $K_{Si-C}$, the corresponding $Q_{Si-C}$ can be estimated by $Q_{Si-C} \approx K_{Si-C}/\nu_0$.[37, 38].

In the literature, the ratio $\frac{A_L}{A_L+A_G}$ is often used to estimate the crystalline fraction of the Si-C phase [12, 33, 34]. Strictly speaking, this approach is only valid if the bond density in the amorphous phase is connected to the Gaussian peak area by the same proportionality constant $Q_{Si-C}$ by which the bond density in the crystalline phase is connected to the Lorentzian peak area. $K_{Si-C}$ for 3C-SiC is reported to be approximately $2.1 \cdot 10^{19}$ cm$^{-2}$ [36, 38], while for amorphous SiC (a-SiC), $K_{Si-C}$ ranges from $2.1 \cdot 10^{19}$ cm$^{-2}$ to $3.6 \cdot 10^{19}$ cm$^{-2}$ [37, 38]. Using $\nu_0 = 790$ cm$^{-1}$ and $\nu_0 = 740$ cm$^{-1}$ for the crystalline and the amorphous Si-C modes, respectively, we estimated $Q_{Si-C} = 2.7 \cdot 10^{16}$ cm$^{-1}$ for c-SiC and $2.8 \cdot 10^{16}$ cm$^{-1} < Q_{Si-C} < 4.9 \cdot 10^{16}$ cm$^{-1}$ for a-SiC. We found that the trends in crystallinity investigated in this work are maintained for all these values of $Q_{Si-C}$. This result means that the formula above cannot be used to calculate the absolute crystallinity but is useful for investigating the crystallinity trends of the Si-C network.

The reflection and transmission (***R&T***) of the samples in the UV-Vis wavelength range (250-1000 nm) were measured using a Varian Cary 500i photo spectrometer with an Ulbricht sphere. If the sample thickness $d$ is known, the absorption coefficient $\alpha$ can be calculated from *R&T* measurements by means of the following relation:

$$\alpha = \frac{\ln\left(\frac{1-R}{T}\right)}{d}. \quad (8)$$

For the determination of the absorption coefficient from FTIR and spectrophotometry, it is crucial to know the exact ***layer thickness***. Therefore, we used three different techniques. (i) To measure the annealed SiC SL thicknesses on Si and on quartz substrate, a Woollam M-2000 spectroscopic ellipsometer was used. (ii)

To determine the annealed SRC SL and ML film thicknesses on Si, scanning electron microscopy (SEM) was applied on cross section images. The instrument used was a Hitachi SU70 with a cold trap, operated at 5-6 kV accelerating voltage and a working distance of 4-6 mm. (iii) The thickness of annealed SRC SL and ML films on quartz was calculated from *R&T* measurements using the software OPTICAL [39].

3 RESULTS AND DISCUSSION

*3.1    GIXRD results*

The GIXRD patterns of the ML stacks with a thickness variation of SRC and SiC sublayers and an annealing of 1000°C for 60 min are shown in Figure 2 (a) and (b), respectively. The results of the grain size analysis are presented in Figure 2 (c) and (d). While the SiC NC grain size is $(2.2 \pm 0.5)$ nm for all sublayer thicknesses, the Si NC grain ranges between $(1.5 \pm 0.5)$ nm and $(3.5 \pm 0.5)$ nm. This is the targeted range for Si NC in SiC, as Löper *et al.* [18] calculated that Si NC in SiC with a diameter of around 2.5 nm show a bandgap of 1.7 eV, which corresponds to the ideal bandgap of the top cell in an all-Si tandem solar cell [2].

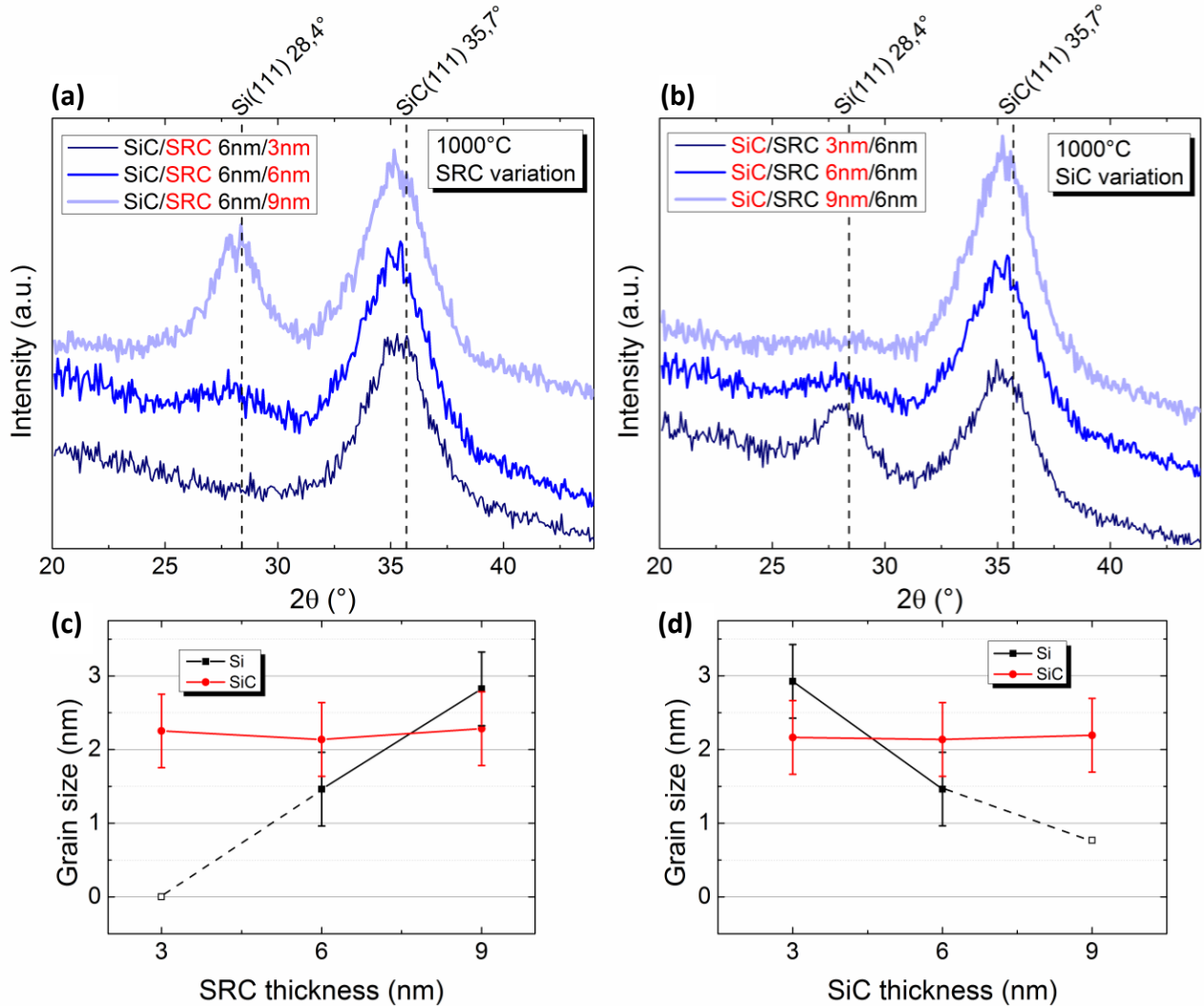

Figure 2: The GIXRD patterns of ML stacks with varying SRC sublayer thickness (a) and varying SiC sublayer thickness (b) were used to calculate the mean grain size of Si and SiC NC. The relation between the sublayer thickness and the grain sizes is shown in (c) for SRC thickness variation and in (d) for SiC thickness variation. The open squares indicate samples for which no definite determination of the Si grain size was possible, and the lines are a guide to the eye.

Closer inspection reveals, that the Si-NC size increases with increasing SRC sublayer thickness, and decreases with increasing SiC barrier layer thickness. The increase in Si NC grain size with SRC sublayer thickness (Figure 2 (c)) is expected from the ML approach. However, for a constant SRC layer thickness of 6 nm at varying SiC barrier layer thicknesses, the ML approach predicts a constant Si NC grain size, which is inconsistent with the trend in Figure 2 (d).

A possible explanation for this behavior is a strong intermixing of the SiC and SRC sublayers, leading to a homogenous distribution of the excess Si throughout the sample. In principle, intermixing seems possible because the estimation of a lower limit for the diffusion coefficient of Si in SiC ($D_{Si}$) during the growth of SiC by carbonization of Si from Cimalla *et al.* [40] leads to the values of $1\cdot10^{-16}$ cm$^{-2}$/s and $2\cdot10^{-15}$ cm$^{-2}$/s at 1000°C and 1100°C, respectively. These values correspond to a diffusion length of 6 nm after 60 min at 1000°C and 60 nm after 30 min at 1100°C, which would be sufficient for complete intermixing even for a ML with 9 nm barriers. Furthermore, intermixing of SiC/SRC ML was observed by many groups using TEM images [11, 14, 41].

We checked the assumption of intermixing experimentally by plotting the Si NC size as a function of the overall Si content in at% ($Si(at\%)$) in Figure 3. The determination of $Si(at\%)$ was performed by first calculating the Si volume fraction $V_{Si}(ML)$ as described by Summonte *et al.* [12]. Then, $V_{Si}(ML)$ was transformed into $Si(at\%)$. The numbers of atoms in a cubic centimeter of Si and SiC ($n_{Si}$, $n_{SiC}$) were as follows:

$$n_{Si} = N_A \cdot \rho_{Si}/m_{Si} = 4.99 \cdot 10^{22} \frac{at}{cm^3} \quad (9)$$

$$n_{SiC} = 2N_A \cdot \rho_{SiC}/m_{SiC} = 9.53 \cdot 10^{22} \frac{at}{cm^3} \quad (10)$$

where $N_A$ is the Avogadro constant and $m_{Si}, \rho_{Si}$ and $m_{SiC}, \rho_{SiC}$ are the molar masses and mass densities of Si and SiC. Weighted with the volume fractions, the numbers of atoms of Si and SiC in the ML ($n_{Si}(ML)$, $n_{SiC}(ML)$) were determined as follows:

$$n_{Si}(ML) = V_{Si}(ML) \cdot n_{Si} \quad (11)$$

$$n_{SiC}(ML) = V_{SiC}(ML) \cdot n_{SiC}. \quad (12)$$

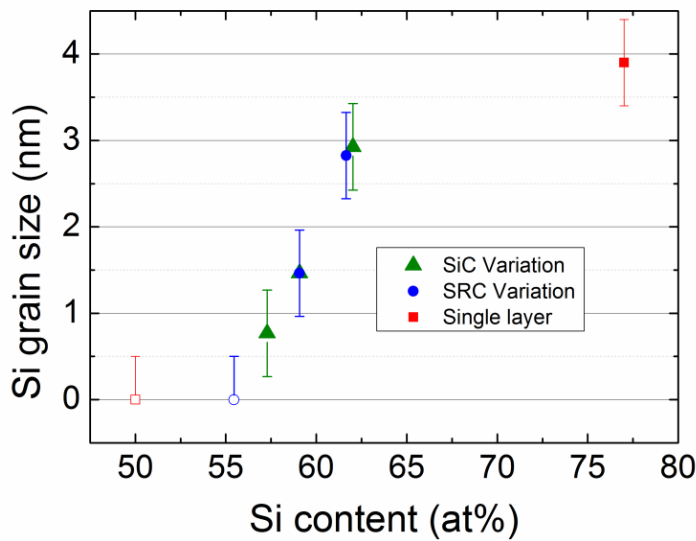

**Figure 3 :** The Si grain size as a function of overall Si content in the ML with SRC sublayer thickness variation (blue), SiC barrier thickness variation (green) and two SL (red). In the samples with 50% and 56% Si content, no crystalline Si was detected by GIXRD.

This led finally to the overall Si in the ML ($Si(at\%)$):

$$Si(at\%) = \frac{n_{Si}(ML) + n_{SiC}(ML)/2}{n_{Si}(ML) + n_{SiC}(ML)}. \quad (13)$$

In addition to the ML already shown in Figure 2, two SL with a composition of $Si_{0.50}C_{0.50}$ and $Si_{0.77}C_{0.23}$ were also taken into account. The continuous increase in Si NC size with overall Si content, regardless of the precursor layer structure, supports the assumption of strong sublayer intermixing.

To obtain further insight into whether the properties of the layers are a function of sublayer thickness variation or of overall Si content or both, a detailed FTIR investigation of the Si-C network was performed.

*3.2    FTIR results*

In Figure 4 (a), a typical fitted $\alpha(\nu)$ spectra, as described in section 2.2, of a ML stack with an overall Si content of 56 at% is depicted. As the peak position of the $L(\nu)$ part correlates very well with the maximum of the peak, we consider $\nu_L$ as the overall peak position $\nu_0$ in all samples. All samples can be fitted properly with a Lorentzian and a Gaussian peak except the sample with the highest overall Si content (77 at%). For this sample, a second Gaussian peak is needed (Figure 4 (b)). We consider this peak to represent a second amorphous Si-C phase that might also be present in the other samples but was too small to be significant and/or too close to the peak position of the first amorphous phase to be observable.

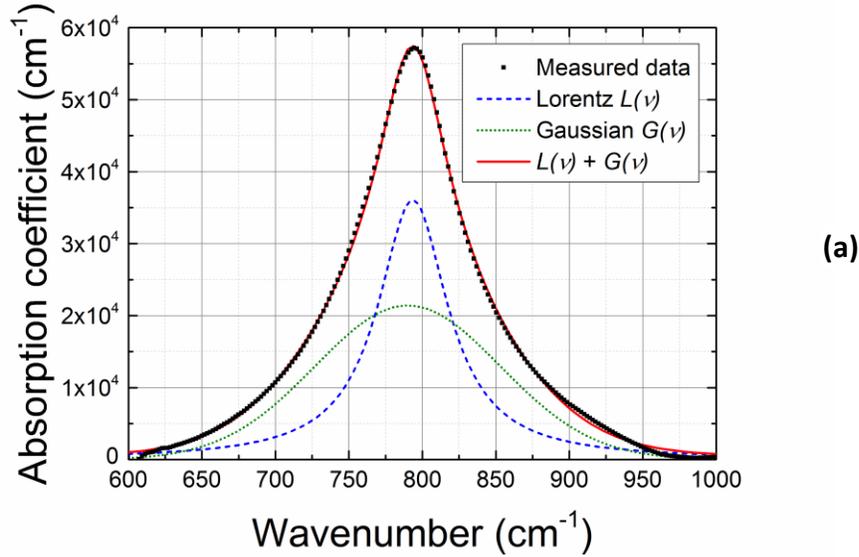

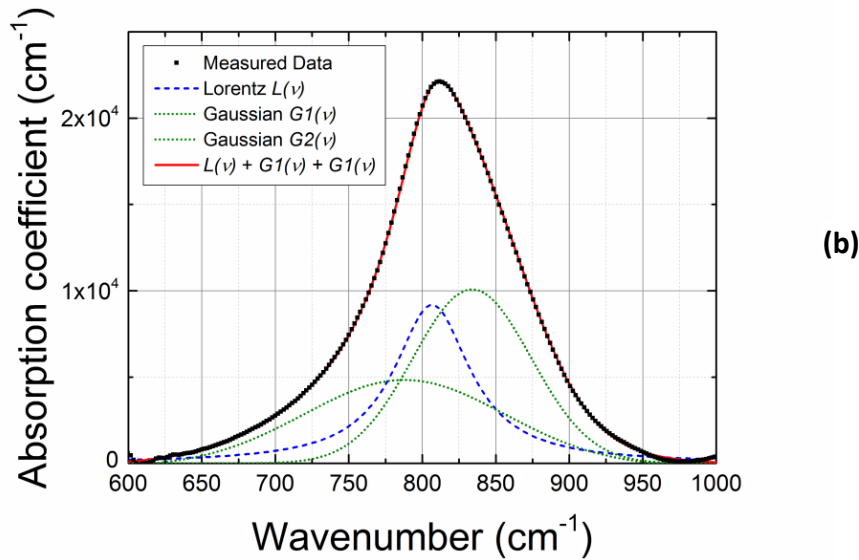

**Figure 4: A typical fitted Si-C mode with a Lorentzian part (blue), a Gaussian part (green) and the overall fit function (red). The sample used in (a) is a ML stack with a well thickness of 3 nm, a barrier thickness of 6 nm and an overall Si content of 56 at%. In (b), the measurement and the fit of the SL with 77 at% Si are shown. It is the sole data set that can only be fitted properly with two Gaussian peaks.**

This FTIR peak fitting was performed for the same samples analyzed by GIXRD in Figure 3, and the results are plotted in Figure 5. The Si-C vibration in stoichiometric $Si_{0.50}C_{0.50}$ is at $\nu_{SiC} = (790.0 \pm 0.1)$ cm$^{-1}$, and $\nu_{SRC} = (806.6 \pm 0.2)$ cm$^{-1}$ for $Si_{0.77}C_{0.23}$. Assuming ML with preserved sublayers of alternating $Si_{0.50}C_{0.50}$ and $Si_{0.77}C_{0.23}$, the Si-C vibrational mode should be a combination of $\nu_{SiC}$ and $\nu_{SRC}$ with varying amplitudes. However, such fitting of the ML FTIR spectra was unsuccessful. Instead, Figure 5 (a) shows that $\nu_0$ increases linearly with Si content. This behavior supports the assumption of sublayer intermixing in the ML. The Si-C crystallinity is shown in Figure 5 (b). Note that for the crystallinity determination of the sample with the highest Si content, both Gaussian peaks shown in Figure 4 (b) were taken into account as amorphous contribution. We observe a decline in the crystalline fraction with increasing Si content. Closer inspection reveals that the decrease in crystalline fraction for increasing SRC thickness (blue points in Figure 5 (b)) is much stronger than for decreasing SiC sublayer thickness (green points in Figure 5 (b)). Assuming ML stacks (whether with or without preserved sublayers) with complete Si/SiC phase separation, we would not expect a dependence of the crystalline fraction on sublayer variation nor on overall Si content. Thus, the observed overall decrease of crystallinity is an indication of incomplete phase separation and of excess Si hindering the Si-C crystallization. The effect of the sublayer variation on the slope of the decrease is interpreted as the influence of the initial ML structure on the phase separation or the degree of intermixing in the Si-C network.

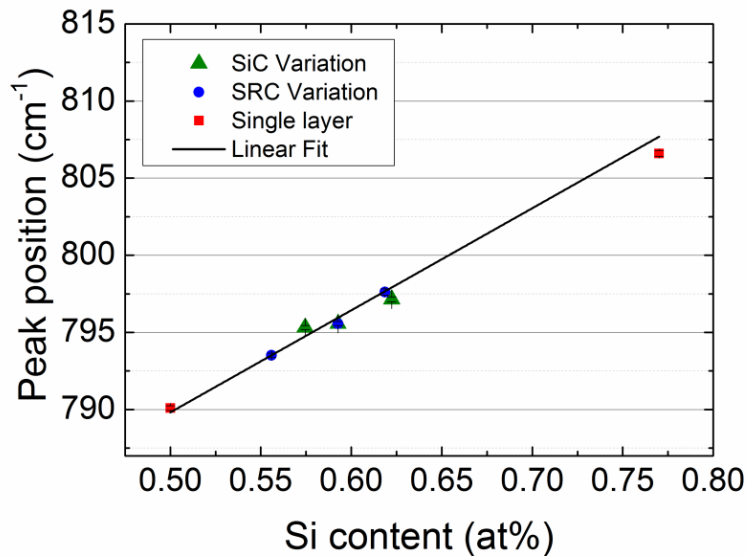

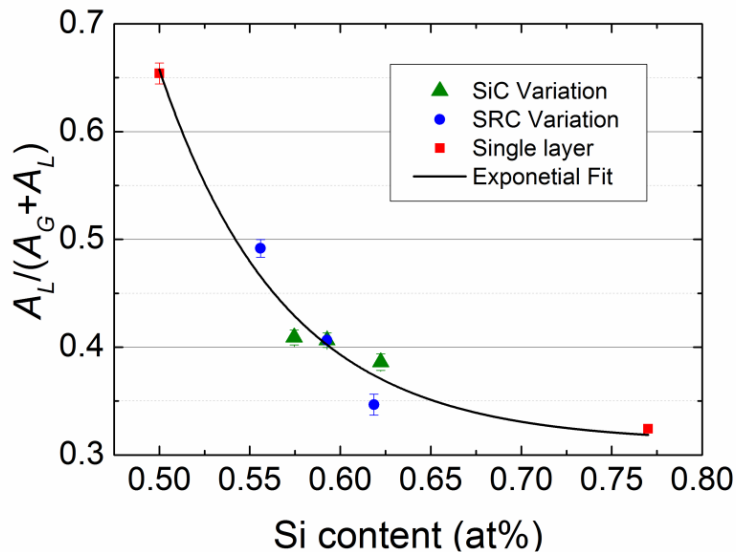

**Figure 5:** The parameters extracted from the fitted FTIR spectra as a function of the overall Si content in the ML with SRC sublayer thickness variation (blue), SiC barrier thickness variation (green) and two SL (red). In (a), the peak position $\nu_0$ of the Lorentzian Si-C vibrational mode is depicted. The contribution of the Lorentzian peak to the Si-C mode indicates the SiC crystallinity and is shown in (b).

The analysis of the Si-C network by FTIR further supports the assumption of sublayer intermixing but suggests that there is still an influence of the sublayers on the binding behavior in the samples. Furthermore, the results show an incomplete phase separation in the ML and a strong suppression of Si-C crystallization by excess Si.

Thus far, we have mainly discussed structural properties of the Si NC layers. Regarding their future use in optoelectronic devices, an investigation of the impact of these structural trends on absorption in the UV-Vis wavelength range is crucial and will be described in the next section.

## 3.3 Spectrophotometry results

For spectrophotometry measurements, two bulk SL with the compositions of the sublayer in the ML ($Si_{0.77}C_{0.23}$ and $Si_{0.50}C_{0.50}$) as upper and lower bound and four ML stacks with varying sublayer thicknesses were used. In Figure 6 (a), the absorption coefficient $\alpha$ for these samples is depicted as a function of photon energy ($E$).

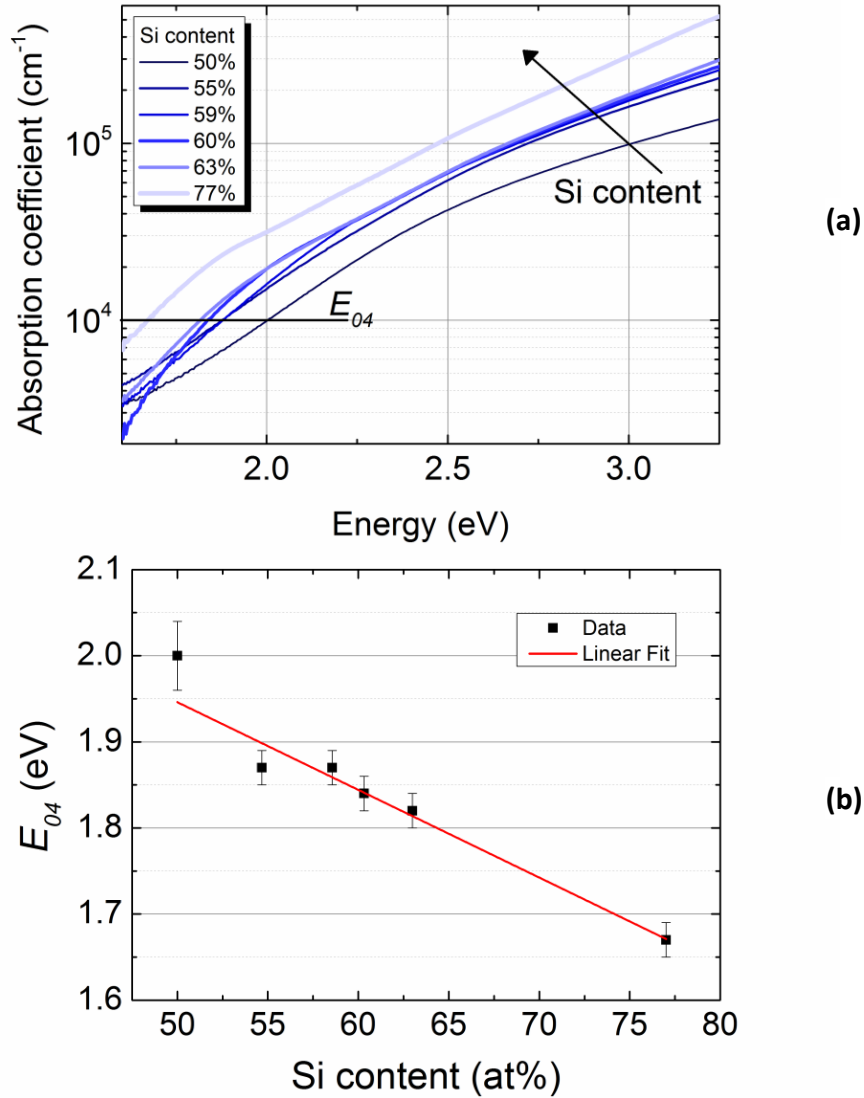

**Figure 6:** In (a), the absorption coefficient $\alpha$ derived from R&T measurements for ML with a different overall Si content is shown. The 50% and the 77% samples are SL. The bandgap estimation $E_{04}$ at $\alpha = 10^4$ cm$^{-1}$ as a function of the Si content is plotted in (b).

Effectively, $\alpha(E)$ of all ML lies between the border cases of bulk SiC and SRC. A reliable determination of the bandgap by evaluating the Tauc plots derived from Figure 6 (a) is not possible, as the Tauc plots show no linear region required for determining the bandgap. This result is not surprising given that this evaluation method is valid only for amorphous semiconductors [42, 43].

Thus, we estimate the bandgap $E_{04}$ by taking the energy at $\alpha = 10^4$ cm$^{-1}$. Figure 6 (b) shows that $E_{04}$ decreases continuously with the Si content.

This behavior could be explained by either an intermixing of the ML that leads to absorption behavior of an Si$_x$C$_{1-x}$ layer or by a decreasing bandgap with increasing Si NC size, as we know from Figure 3 that the Si content correlates with the Si grain size.

*3.4    Degree of intermixing*

The previous results showed clearly that strong intermixing of the sublayers in the ML stacks occurs during high-temperature annealing. To finally verify whether there is a difference in crystallization behavior in SL and ML, we took a ML stack and a SL of the same overall Si content, both annealed at 1100°C for 30 min, and compared their XRD patterns and their trend of *α(E)*.

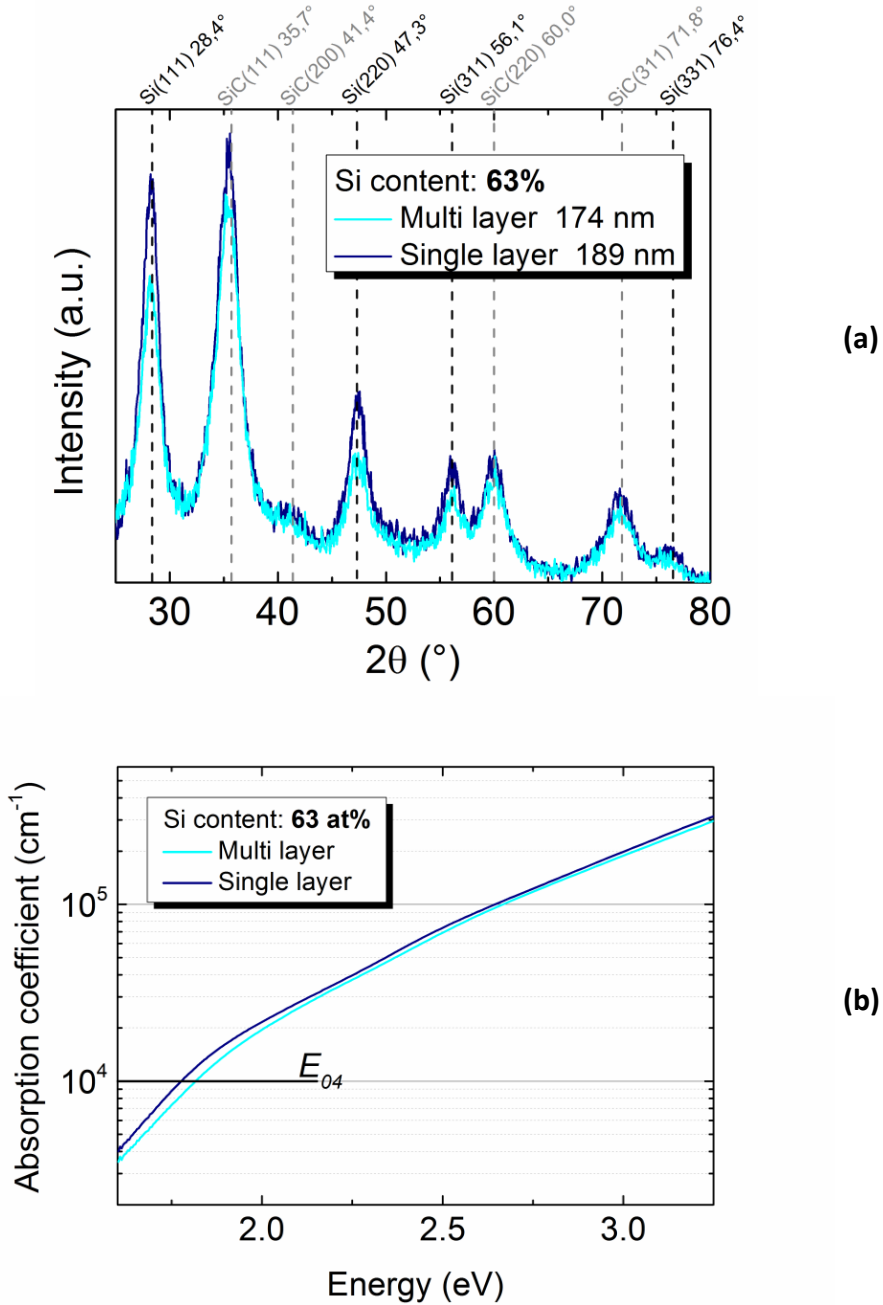

Figure 7: The comparison of a ML stack and a SL with the same overall Si content: In (a) their XRD patterns and in (b) their absorption coefficient as a function of energy is shown.

The XRD pattern in Figure 7 (a) shows a higher overall intensity of the SL compared with the ML, which is most likely due to the small differences in sample thickness and allows no final conclusion about deviations in crystallization behavior. However, the relation of the SiC/Si peak height for a specific crystal plane and the NC sizes contain such information. A grain size increase of NC in the SL compared with the ML is observed but allows no conclusions, as the differences lie within the grain size error of ± 0.5 nm. The evaluation of the peak heights shows that SiC(111)/Si(111) = 1.06 ± 0.05 and 1.27 ± 0.07 in SL and ML, respectively. The SiC/Si peak height ratios resulting from the (220) and (311) crystal planes are also higher for ML than for SL but are not specified here, as the difference is not significant due to the large fitting error for smaller peaks.

The simplest explanation for these results could be a slight deviation from the nominal composition in the samples. However, provided that the calculated Si content of 63% (as described in section 3.1) corresponds to the real Si content in the samples, this peak analysis leads to the assumption that Si-crystallization in SL is favored, which is somewhat surprising, as the local Si concentration in the SRC sublayer in the ML stack is much higher than the local Si concentration in the SL. This result suggests that Si diffusion occurs at lower temperatures than Si crystallization and thus is not limited by the question of whether two Si atoms were deposited close to each other.

We suggest that the reduced Si crystallization in the ML is a hint toward incomplete intermixing of the sublayers. We know from GIXRD measurements not shown here that SiC crystallization begins at lower temperatures than Si crystallization. The onset of crystallization in the SiC sublayers causes stress in the layer because the lattice constants *a* of SiC and Si are quite different ($a_{SiC}$ = 4.3596 Å, $a_{Si}$ = 5.431 Å). We consider this stress to hinder Si crystallization, as predicted by the calculations of Summonte *et al.* [12]. Thus, the weaker Si crystallization in ML can be attributed to stronger SiC crystallization due to the incomplete intermixing of SRC and SiC sublayers. It is worth noting that the same argument can be used to explain the trends in Figure 2(d). Thicker SiC sublayers are likely to impose greater stress on SRC sublayers of constant thickness as the SiC sublayers crystallize, slowing the growth of Si NC in ML with thicker SiC sublayers. However, stress is insufficient to explain the behavior of the parameters derived from FTIR measurements.

The value $\alpha(E)$ shows the same trend for ML and for SL. The difference in $E_{04}$ between SL and ML exceeds the experimental error of the absorption measurement, but it is unclear whether this result is due to incomplete intermixing of the ML or a slight deviation from the nominal composition.

3 CONCLUSIONS

We studied ML stacks consisting of SRC/SiC sublayers with a view to achieving Si NC size control and observed strong sublayer intermixing even for the thickest barrier layer of 9 nm (as deposited). The Si NC grain size is a function of the overall Si content in the ML and not only of the SRC well thickness. The IR absorption of the Si-C network in the ML cannot be modeled as a superposition of two sublayer modes but was found to be a continuous function of the Si content, confirming interdiffusion. The trend of Si-C crystalline fraction as a function of sublayer thicknesses and overall Si was interpreted in terms of incomplete Si/SiC phase separation and of excess Si hindering the Si-C crystallization. However, it was shown that there is still an influence of the sublayers on the binding behavior in the samples. Only the optical properties cannot be clearly ascribed to sublayer intermixing, as the decreasing trend of $E_{04}$ with increasing Si-content can be an effect of either quantum confinement or intermixing. Furthermore, we found hints of greater stress in ML compared with SL by direct comparison of SL and ML with the same overall Si content.

Based on these results, we propose the use of SL with varying Si content to control the Si NC grain size instead of using ML with varying SRC thickness. The examined Si NC mean grain sizes between 1 nm and 4 nm lie in the targeted range for Si NC in SiC as a top cell absorber in an all-Si tandem solar cell [2]. It is true that in SL, the control of crystal size, size distribution and interdot distance is limited. However, the formation of large grains (> 10 nm) can be excluded for all samples by means of GIXRD measurements. Furthermore, both the interdot distance and the grain size distribution are less crucial for Si NC in SiC than for Si NC in $SiO_2$ [8, 18]. Consequently, precise control of Si NC size and separation is not required, and as this study and other works [11, 12, 14] show that control over both is extremely difficult to obtain with the ML approach in the Si NC/SiC system, it follows that SRC SL might be more effective for optimising the optoelectronic properties of SiC with embedded Si NC for Si NC-based devices. The thickness of SL films need not be as precisely controlled as the thickness of sublayers in an ML, enabling faster film deposition and significantly reducing the cost of producing Si NC/SiC devices.

4 ACKNOWLEDGEMENTS

The authors wish to thank Manuel Moser, Antonio Leimenstoll, Felix Schätzle and Mathias Rumpel for help with sample processing and Gina Kraft for *R&T* measurements. The research leading to these results has received funding from the European Community's Seventh Framework Programme (FP7/2007-2013) under grant agreement n°: 245977 under the project title NASCEnT.

*Published in:* *Applied Surface Science, vol. 351, p. 550-557, 2015.*
*DOI:* 10.1016/j.apsusc.2015.05.153